\magnification=1200
\nopagenumbers
\parskip=10pt
\vskip .7truein
\centerline {\bf{ Spurious Effects in Perturbative Calculations}}
\vskip 1.0truein
\centerline{ M.Horta\c csu and B.C.L\" utf\" uo\u glu}
\vskip .3truein
\centerline { Physics Department, Faculty of Science and Letters}
\vskip .1truein
\centerline { I.T.U., 80626, Maslak, Istanbul, Turkey}
\vskip 1.6truein
\noindent
Abstract: We show spurious effects in perturbative calculations due to
different orderings of inhomogenous terms while computing corrections to
Green functions for two different metrics.  These effects are not carried
over to physically measurable quantities like the renormalized value of
the vacuum expectation value of the stress-energy tensor.

\def\kutu{{\rlap{$\sqcup$}\sqcap}}
\baselineskip=18pt
\footline={\centerline {\folio}}
\pageno=1
\vfill\eject
\vskip  .1truein
\noindent
{\bf{Introduction}}

\noindent
Finding logarithmic behaviour of correlation functions instead of pure power
law has always been interesting in quantum field theory.  QCD  differs from  
a free theory at the asymptotic region only by its logarithmic corrections.
The presence of these corrections differentiate a physically interesting 
theory from a trivial model.  Another effect is the possibility that
logarithmic corrections in perturbation theory may sum up to an anamolous
power law, as seen in the Thirring model [1].  Keeping these examples
in mind, it is always get exciting to encounter logarithmic behaviour while
calculating Green functions.  That is why one may be curious whether
encountering such terms always points  to important physical phenomena.

Here we want to study the interaction of a gravitational field described by a certain
metric with a scalar field.  We do not know how to quantize the gravitational
field, though.  This fact necessitates the use of semi-classical methods .
These methods treat the gravitational field classically and couples it to
the scalar field only by writing the d'Alembertian in the background of
that metric [2].  This operation reduces the full field theoretical problem
to an external field calculation.  We study the n-point functions of the
problem and try to deduce information about the full theory
from these quantities.  Out of these n-point
functions, the two point function, which is the Green function of the 
d'Alembertian,
is the most important one, since by differentiating it we can get the vacuum
expectation value, VEV, of the
stress-energy tensor, $T_{\mu \nu}$.

To be able to calculate renormalized value of the VEV of the stress-energy
tensor, $<T_{\mu \nu}>_{ren}$, we have to find an algorithm
to regulate the divergences of the two-point function, $G_F$, at the
coincidence
limit. The initial step in this programme is to study the singularity
structure of $G_F$ for the particular model studied.
We know that the singularity  at the coincidence limit of the Green
function for a free scalar field in flat space is  quadratic.
At this point one is confronted 
with a very important theorem [3] concerning the singularity behaviour of
Green functions in different backgrounds.
This theorem states that if a metric is flat at any region in space-time,
the singularities of the Green function in the background of this metric
should exhibit  Hadamard behaviour [4], which is at
worst quadratic.  Although we may by-pass the results of this theorem by
studying metrics that are not $C^{\infty}$, still  it is very improbable
for Green functions of scalar particles
in the
background  metrics  that are flat at any
region in space-time to have singularities that are worse than quadratic.

The theorem quoted above makes us suspicious
of any extra logarithmic terms we encounter while calculating the
Green functions in the background of different metrics, especially if 
the calculation is performed perturbatively.  We want to see if such 
terms are genuine or the result of the different regularization
procedures used, arising when we make our expressions unnecessarily
singular.  Whether such effects are genuine or not can be checked by
comparing perturbative results with the exact ones, when available, and 
studying different ways of grouping terms with more or less singular 
behaviour.

Here we are going to continue  our investigation of spurious effects in
Green function calculations using different metrics [5].  We will see that
while we take alternative routes in solving inhomogenous partial differential
equations results in solutions with different singularity structure for the
Green's function, this difference cancels out in the calculation of physical
quantities like the VEV of the stress-energy tensor,
after it is renormalized.

Our model is an impulsive gravitational plane wave solution of Hogan [6].
$$ds^2=-2\cos^2{ aU} |dz+\Theta(U) {{\tan{aU}}\over {a}} {\overline{H}}
({\overline{z}} ) d{\overline{z}}|^2+2dUdV ,\eqno {1}$$
with the only one non-vanishing Ricci tensor component,
$$R_{44} = -2a^2 \Theta (U), \eqno {2}$$
and only one non-vanishing Weyl tensor component ,
$$\Psi_{4} = H(z) \delta (U). \eqno {3} $$
Here $\Theta$ is the Heavyside unit step function,$\delta$ is the Dirac
delta function, $a$ is a constant
and $H$ is the $z$ derivative of a smooth arbitrary function of $z$,
$H(z)={{dG}\over {dz}}$.   U and V are null coordinates,
$z={1\over {\sqrt{2}}}(x+iy)$.
This metric is flat for $U<0$ and conformally flat for $U>0$.
This metric is similar to the metric found by Nutku  earlier [7] .

We have two reasons for studying  this model .  First, we have the exact
result to compare with the perturbative one,  in one of the two
cases studied. Second, we know that the $<T_{\mu \nu}>_{ren}$ should be
null $^{/8,9}$,
a result we have to obtain at the end.

We see that this metric has a dimensional parameter $a$.  We have seen that
the presence of a dimensional parameter in the metric may result in
non-standard ( non-Hadamard) behaviour in the Green function , when
this quantity is calculated perturbatively $^{/10}$.
Here we find that this change in the singularity structure occurs if we group
our terms in a certain manner in the perturbation expansion even in
the first order calculation.  It is absent
for the exact calculation .   It is amusing, however, to see that these terms are absent
if we group the inhomogenous terms in a different manner.
This fact shows that the change in the singularity in the former calculation
is due to the wrong choice of the parameters, which results in severer
singularities, and, when regulated in the Schwinger formalism, ends
up with terms which are worse than the other case by logarithms.
We find that this is another method for generating 
logarithmic singularities in the coincidence
limit for the Green functions at will .  A similar phenomenon was shown
to exist in  [10], by taking the homogenous solutions into
account.
These singularities do not
survive when the VEV of the renormalized stress-energy tensor, a measurable
physical quantity, is  calculated.

We  first solve the Green function, $G_F$ for a special $H(z)$ used
in the metric given by eq. [1] exactly and point to the 
Hamadard  behaviour of this expression.  Then we get the perturbative
solution in two ways, and show two different results obtained for $G_F$, only
one of them
in the Hadamard form.  In the third section we use a more complicated 
form of $H(z)$ , where we could not obtain
the exact result and show the same conflicting behaviour when $G_F$ is
calculated perturbatively.  At the end we note that when the VEV of the
stress-energy tensor is
computed the different expressions for $G_F$ give the same result
for $<T_{\mu \nu}>_{ren}$. All through our work we  use  conformal coupling.
\vskip.4truein
\noindent
2.1{  \bf{ Exact Calculation for the First Metric}}

\noindent
We choose $G=az$ for the arbitrary function in the metric given above, which
gives the following  expression for the d'Alembertian operator written 
in the background of this metric
$$\kutu= 
2{{\partial^2}\over{\partial U \partial V}}
-2a\tan(2aU){{\partial}\over { \partial V}}
+{{4}\over{\cos^2(2aU)}}\left(\sin(2aU)-1\right) {{\partial^2}
\over{\partial x^2}} $$
$$-{{4}\over{\cos^2(2aU)}} \left(\sin(2aU)+1\right) 
{{\partial^2}\over{\partial y^2}} . \eqno {4}$$
If we write
$$\kutu f(x,y,U,V)=0, \eqno {5}$$
we find
$$f={{e^{iRV}e^{ik_1x}e^{ik_2y}}\over {\cos^{1/2} (2aU)}}
e^{{{k_1^2-k_2^2}\over {iRa\cos(2aU)}}}
e^{-{{k_1^2+k_2^2}\over{iRa}} \tan(2aU)}. \eqno {6}$$
Here $k_1,k_2,R$ are the Fourier modes that will be integrated over in
the Green's function calculation.  These integrations are performed easily 
and we get 
$$G_F= -{{a\left(\Theta(U-U')-\Theta(U'-U)\right)}\over
{8\sqrt{2}i\pi^2 \sin{2a(U-U')}\left( (V-V')
-a{{(x-x')^2}\over{4\Delta^2}} -a{{(y-y)^2}\over{4\Delta^2_1}} \right)}},
\eqno {7}$$
where
$$ \Delta^2=\left(-{{1}\over{\cos(2aU)}}+\tan(2aU)+{{1}\over
{\cos(2aU')}}-\tan(2aU')\right) ,\eqno {8} $$
$$ \Delta^2_1= \left( {{1}\over {\cos(2aU)}}+\tan(2aU)-{{1}\over
{\cos(2aU')}}-\tan(2aU') \right) . \eqno {9} $$
We see easily that in the coincidence limit this function has a quadratic 
divergence, same as in the flat metric.
\vskip.4truein
\noindent
{2.2{\bf{ Perturbative Calculation for the first metric}}

\noindent
We expand the operator $\kutu$ in powers of $a$.
If we write
$$\kutu= L_0+aL_1+..., \eqno {10}$$
we find
$$L_0= 2{{\partial ^2}\over {\partial U \partial V}}
-{{\partial ^2}\over {\partial x^2}}-{{\partial ^2}\over{ \partial y^2}},
\eqno {11}$$
$$L_1= 2U\left( {{\partial ^2}\over {\partial x^2}}-{{\partial ^2}\over 
{\partial y^2}}\right). \eqno {12}$$
We also expand both the eigenfunction and the eigenvalue of the equation
$$ \kutu \phi= \lambda \phi \eqno {13}$$
in powers of $a$ as $ \phi= \phi^{0} + a \phi^{1}+... $ and 
$\lambda= \lambda^{0}+\lambda^{1}+...$.
We find 
$$\phi^{0}={{1}\over {(2\pi)^2(2|R|)^{{1}\over{2}}}}
e^{i\left({{K}\over {2R}} U+RV+k_1x+k_2y \right)},
\eqno {14}$$
$$\lambda^{0}=K-k^2_1-k^2_2,\eqno {15} $$
$$\lambda^{1}=(\phi^{0},L_1 \phi^{0} )=0.\eqno {16}$$
Here $k_1,k_2,K,R$ are the different modes we have to integrate over to find 
the Green function $G_F$. $\phi^{1}$ satisfies the equation
$$(L_0-\lambda_0)\phi^{1}+L_1 \phi^{0} =0 .\eqno {17} $$
The form of this equation suggests the ansatz
$$ \phi^{1}= \phi^{0}(g(U,z)+h(U,\overline{z})) \eqno {18}$$
where $z={{1}\over {\sqrt{2}}}(x+iy)$.
This ansatz yields the following equation for the unknown functions
$g$ and $h$:
$$iR\left({{\partial g}\over {\partial U}} 
+ {{\partial h}\over {\partial U}}\right)
-{{i}\over {\sqrt{2}}}(k_1+ik_2){{\partial g}\over {\partial z}} 
-{{1}\over{\sqrt{2}}}(k_1-ik_2)
{{\partial h}\over {\partial {\overline {z}}}}
+(k_1^2+k_2^2)U=0 .\eqno {19}$$
We seperate this equation into two parts, one for  
a function of $z$ and $U$,
and the other a function of $\overline {z}$ and $U$ .
One choice for this  decomposition is taking
$$iR {{\partial g_1}\over {\partial U}}-{{1}\over {\sqrt{2}}}(k_1+ik_2)
{{\partial g_1}\over {\partial z}}
+{{1}\over{2}} (k_1^2+k_2^2) U=0, \eqno {20}$$ 
and
$$iR {{\partial h_1}\over {\partial U}}-{{i}\over {\sqrt{2}}}(k_1-ik_2)
{{\partial h_1}\over { \partial {\overline {z}}}}
+{{1}\over {2}} ( k_1^2+k_2^2) U=0 . \eqno {21}$$
At this point a clearification is in order.  
Since the inhomogeneous term is a constant, as
far as $z$ and ${\overline {z}}$ are concerned, there is ambiguity how it is
shared
among the two equations.  Here we designate by $g_1, h_1$, the particular
choice for the decomposition given above.

Once this seperation is made, the integration is immediate.  We find
$$g_1=-{{(k_1^2+k_2^2)}\over {i\sqrt {2}(k_1+ik_2)}} Uz-iR {{(k_1^2+k_2^2)}
\over {2(k_1+ik_2)^2}} z^2 , \eqno {22}$$
$$h_1={{i(k_1^2+k_2^2)}\over {\sqrt{2}(k_1-ik_2)}} U{\overline {z}}
-iR {{(k_1^2+k_2^2)}\over {2(k_1-ik_2)^2}}
{\overline {z}}^2 . \eqno {23}$$
To calculate the Green function $G_F$ we have to sum over all the modes,
$$G_F=-\sum _{\lambda} {{\phi \phi^{*}}\over {\lambda}} ,\eqno {24}$$
which reduces in our first order calculation to 
$$G_F^{(1)}= -\sum  {{\phi_{0} \phi_{0}^{*}[g_1+g_1^{*}+h_1+h_1^{*}]}
\over {\lambda_{0}}} .\eqno {25}$$
When written explicitly, we get
$$G_F^{(1)}=
-\int^{\infty}_{-\infty} dR \int^{\infty}_{-\infty} dK
\int^{\infty}_{-\infty} dk_1 \int^{\infty}_{-\infty} dk_2 \times $$
$$\times {{e^{ \left( i[R(V-V')+{{K}\over {2R}}(U-U')+k_1(x-x')+k_2(y-y')]
\right)}}
\over { (2\pi)^4(2|R|)(K-k^2_1-k^2_2)}}(F(\chi)+F^{*}(\chi')), \eqno {26} $$
where
$$ F= -i{{(k_1^2-k_2^2)}\over {k_1^2+k_2^2}}\left((k_1x+k_2y)U+
{{R}\over{2(k_1^2+k_2^2)}}
[ (k_1^2-k_2^2)(x^2-y^2)+4k_1k_2xy]   \right), 
\eqno {27}$$
and $\chi$ is the generic name for the four variables $U,V,x,y$.
We use the Schwinger representation to write
$${{1}\over {K-k_1^2-k_2^2}}=i \int _{0}^{\infty} d\alpha
\exp (-i(K-k_1^2-k_2^2)\alpha-\alpha \delta)  , \eqno {28}$$
$${{1}\over{(k_1^2+k_2^2)^2}}=-\int_{0}^{\infty} d\beta \beta \exp  
{(-i(k_1^2+k_2^2)\beta -\gamma \beta)} \eqno {29}$$
where $\delta$ and $\gamma$ are infinitesimal positive and real quantities.
We perform the integrals in the usual manner.  The end result is written 
using an infrared parameter $m$ and the zeroth order Hankel function
$ H_0^{(2)}$
as
$$G_F={{1}\over{4\sqrt{2} (2\pi)^2}}\left( {-{A}\over {B\Delta}} 
+{{i\pi A}\over{B^2}} H_0^{(2)} (2m \sqrt {\Delta}) + ...\right) ,
\eqno {30}$$
where $... $ contains terms with the same singularity behaviour as the first
two, i.e. terms with quadratic divergence and quadratic divergence times a
logarithmic divergence in the coincidence limit. 
The logarithmic divergence is given by the $H_0^{(2)}$ term 
which goes to a logarithm as its argument goes to zero when the infrared
cut-off is removed,
$$H_0^{(2)}\propto \log {\Delta} +\log{2m} ,\eqno {31} $$
and we discard the $\log {2m} $ term.
In the above expression 
$$A= (x-x')(Ux-U'x')-(y-y')(Uy-U'y'), \eqno {32} $$
$$ B=(x-x')^2+(y-y')^2, \eqno {33}$$ 
$$\Delta=2(U-U')(V-V')-(x-x')^2-(y-y')^2 . \eqno {34}$$
In this expression we see that the expected Hadamard behaviour, i.e. the
worse divergence being only quadratic, is modified by a logarithm.

It is amusing to note that this behaviour, which is not reflected to the
exact solution,
is an artefact of the choice we used in separating our eq. [19] into the
holomorphic and antiholomorphic parts.
Another choice is separating eq. [19] as
$$ iR {{\partial g_2} \over {\partial U}} - {{i}\over{\sqrt{2}}}(k_1+ik_2)
{{\partial g_2}\over {\partial z}} 
+{{(k_1+ik_2)^2}\over {4}}U=0  , \eqno {35} $$
$$ iR {{\partial h_2} \over {\partial U}}-{{1}\over {\sqrt {2}}}(k_1-ik_2)
{{\partial h_2} \over {\partial {\overline {z}}}}
+{{(k_1-ik_2)^2}\over {4}}U=0 . \eqno {36}$$

The integrations of these equations give immediately
$$g_2={{-i}\over {2\sqrt{2}}}(k_1+ik_2) Uz-iR{{z^2}\over {4}}, \eqno {37}$$
$$h_2={{-i}\over {2\sqrt{2}}}(k_1-ik_2)U{\overline{z}}
-iR{{{\overline{z}}}\over {4}} . \eqno {38}$$
Note that with the latter choice for decomposition of the equation, we have
reduced the powers of $(k_1+ik_2),(k_1-ik_2)$ in the denominator.
Since we integrate these expressions from minus infinity to plus infinity,
terms in the denominator vanish in this range and we use the Schwinger
prescription given above to regulate them.  
The calculation of $G_F$ with less severe divergences is much simpler now.  
The Green function 
integration is straightforward.  For $U>U'$, it reads
$$G_F={{1}\over {4\sqrt {2} (2\pi)^2}} \times $$
$$\times \left({{C_1[(x-x')(Ux-U'x')-(y-y')(Uy-U'y')]
+C_2[(u-u')(x^2-y^2-x^{'2}+y^{'2})]} 
\over {\Delta^2}} \right) \eqno {39} $$
where $C_1, C_2$ are two constants.   This singularity has the same
singularity behaviour as the exact solution at the coincidence limit.

\vskip .4truein
\noindent
{\bf{3. Second metric}}

\noindent
Here we  show the same thing with a different function $G(z)$ used to  
specify the metric in eq. [1] explicitly, to illustrate that the phenomena
we find is not special to only one choice of the trial function $H$
.  We take the next simplest form,
$G={{az^2}\over {2}} $.  Then the metric reads, in the region where
it is not flat,
$$ds^2=2dUdV-2\left[dzd{\overline{z}}\left( \cos^2(aU)
+a^2z{\overline {z}}\sin^2(aU) \right)
+{{a}\over{2}} sin(2aU)\left( z(d{\overline {z}})^2
+{\overline {z}}(dz)^2 \right) \right]. \eqno {40}$$ 
We find 
$$\kutu = -{{2a\cos (aU)  \sin (aU) (1+z{\overline {z}} )}\over {B}}
{{\partial } \over {\partial V}} +2 {{\partial ^2}\over {\partial U
\partial V}}
+{{2a \cos ^3 (aU) \sin (aU) }\over {B^3}}
\left( {{\partial }\over {\partial z}}+{{\partial }\over 
{\partial {\overline {z}}}} \right)
$$
$$+{{2a\cos (aU) \sin (aU)} \over {B^2}}
\left( z{{\partial ^2}\over {\partial z^2}}
+{\overline {z}} {{\partial ^2} \over {\partial {\overline {z}} }} \right)
-{{2a^2 \cos^2(aU) \sin^2 (aU)} \over {B^2}} \left( z{{\partial}
\over {\partial z}}
+{\overline {z}} {{\partial} \over {\partial {\overline {z}}}}\right)    $$
$$ -2{{\cos^2(aU)+a^2 z{\overline {z}} \sin^2 (aU)}\over {B^3}}
{{\partial ^2} \over {\partial z \partial {\overline {z}} }}. \eqno {41} $$
Here $B= \cos^2(aU)-a^2z{\overline{z}} \sin^2(aU) $.
We could not solve the Green function of this operator exactly; so, we do not
know the exact singularity structure of  it.  
We compare, however, the singularity structure of the two expressions 
we obtain by grouping the inhomogeneous terms differently.
Just as in the first example, one way results in a function with Hadamard
behaviour,
the other gives rise to a term which is modified by logarithmic corrections.

We expand the operator $\kutu$ in powers of $a$.  At the first nontrivial
order
we get
$$ \kutu\approx 2\left({{\partial^2}\over {\partial U \partial V}}
-{{\partial^2}\over {\partial z \partial {\overline{z}}}} \right)
+2a^2U\left( -{{\partial}\over{ \partial V}}+{{\partial }\over {\partial z}}
+{{\partial} \over {\partial {\overline {z}}}} 
+z{{\partial^2}\over{\partial {\overline {z}} ^2}}
+{\overline{z}} {{\partial^2}\over {\partial z^2}} \right). \eqno {42}$$
We separate $\kutu$ into two parts $\kutu\approx L_0+a^2L_1$ and
expand both the eigenvalue $\lambda$ and the eigenfunction $\phi$ 
of the equation 
$$\kutu \phi= \lambda \phi \eqno {43}$$
in the same manner.  The zeroth-order and  first-order eigenvalues 
and zeroth-order eigenfunction are as
given in eq.s (14)-  (16).  
We get the equivalent of eq. [17] , with new $L_1$
for $\phi_1$.  The structure of this equation again suggests the ansatz of
eq. [18],
$\phi_1=\phi_0[g(U,z)+h(U,{\overline {z}})] $ which gives rise to 
the equation
$$iR{{\partial }\over {\partial U}} (g+h)-\left({{ik_1-k_2 }\over 
{\sqrt {2}}}\right){{\partial g}\over {\partial z}}
-\left({{ik_1+k_2}\over {\sqrt {2}}} \right){{\partial h}\over
{\partial {\overline {z}}}}$$
$$+U\left[ z \left( {{ik_1-k_2}\over {\sqrt{2}}} \right)^2
+{\overline {z}} \left( {{ik_1+k_2}\over {\sqrt{2}}}\right)^2-\sqrt {2}
ik_1-iR\right]=0. \eqno {44}$$
Here we get two different results depending on how we separate this equation
into two equations for the two unknown functions.
One choice is to write
$$iR{{\partial g_1}\over {\partial U}}-\left( {{ik_1-k_2}\over {\sqrt {2}}}
\right)
{{\partial g_1}\over {\partial z}}+{{Uz}\over {2}} (ik_1-k_2)^2
-{{iRU}\over {2}}
+{{iUk_1}\over {\sqrt {2}}}=0, \eqno {45}$$
$$iR{{\partial h_1}\over {\partial U}}-\left( {{ik_1+k_2}
\over {\sqrt {2}}} \right)
{{\partial h_1}\over {\partial {\overline {z}}}}
+{{U{\overline {z}}}\over {2}} (ik_1+k_2)^2
-{{iRU}\over {2}} +{{iUk_1}\over {\sqrt {2}}}=0. \eqno {46} $$
These equations are integrated to get
$$g_1={{iRz^3}\over {6}}+\left[{{(ik_1+k_2)^2 R^2}\over {2(k_1^2+k_2^2)^2}}
+{{\sqrt{2}
(ik_1+k_2)^2k_1R}\over {2(k_1^2+k_2^2)^2}}-{
{\sqrt{2}}\over {4}} U(ik_1-k_2) \right]z^2
$$
$$+{{U(iR-i\sqrt{2} k_1)(ik_1+k_2)}\over {k_1^2+k_2^2}} z \eqno {47}$$
and the corresponding expression for $h_1$ where the $z$ is replaced by
${\overline {z}}$
and $ik_1+k_2$ goes into $ik_1-k_2$ and vice versa.
If we use this solution to obtain the Green function, we obtain functions
with quadratic singularity at the coincidence point  as well as functions
whose singularities are modified by a logarithmic term.
It is also amusing that the logarithmic behaviour comes out only when we
have an ambiguity in seperating the equation.  The first , second  ,
fourth and the fifth terms have only the quadratic Hadamard singularity,
whereas
the third and the sixth terms , in this form,  give rise to logarithms.
The details of this
calculation can be found in reference [11].  We want to state only that
when we calculate the Green function for the term that reads 
$$ 2k_1R{{(ik_1+k_2)}\over {(k_1^2+k_2^2)^2}} z^2 \eqno {48}$$ 
plus the ${\overline {z}}$ part 
, we get 
$${{(U-U')(x-x')}\over {((x-x')^2+(y-y')^2)^2}} (x^2+x^{'2}-y^2+y^{'2}
+2xy+2x'y') H_0^{(2)} (2m \Delta) \eqno {49}$$ 
plus terms with the same singularity structure in addition to terms with
only quadratic singularity structure.  $\Delta$ in this expression was
defined in eq. [34].  We can modify the third and the sixth term in
eq. [47] to reduce the power of
the terms in the denominator, though, and
write the differential equations as
$$R{{ \partial g_2}\over  {\partial U}} -{{(k_1+ik_2)}\over {\sqrt {2}}}
{{\partial g_2}\over {\partial z}}-{{Uz}\over {2i}} (k_1+ik_2)^2 
-{{RU}\over {2}}
-{{U}\over {\sqrt{2}}}(k_1+ik_2) =0, \eqno {50}$$
$$ R{{\partial h_2}\over {\partial U}} -{{(k_1-ik_2)}\over {\sqrt {2}}}
{{\partial h_2}\over {\partial {\overline {z}}}}
-{{U{\overline {z}} }\over {2i}}
(k_1-ik_2)^2-{{RU}\over {2}}-{{U}\over {\sqrt {2}}}(k_1-ik_2)=0 .\eqno {51}$$
In this expression  we keep most of the terms same as those given in eq.[45] 
and change only one term , the only term which gave the logarithmic
correction.

The solution of the above equations reads
$$g_2={{iRz^3}\over {6}}+ iz^2 \left( {{U(k_1+ik_2)} \over {\sqrt {2}}}-
{{R^2}\over {2(k_1+ik_2)^2}} +{{\sqrt {2} i R}\over {2(k_1+ik_2)}} \right)
-zU\left(1+{{iR}\over {\sqrt {2} (k_1+ik_2)}}\right) \eqno {52}$$
and the similar expression for $h_2(U, {\overline {z}}) $.
We check the behaviour in the coincidence for parts of $G_F$ that are
replaced in the new expression.  The terms that are changed are the fourth 
and the fifth terms in eq. [52].  We can show that these new terms give
rise to only to quadratic divergence in $G_F$  .The calculation of the
fifth term gives rise to 
$$ {{2(u-u')}\over {((x-x')^2+(y-y')^2) \Delta}} 
\left( (x-x')(x^2+x^{'2}-y^2-y^{'2})+2(y-y')(xy+x'y') \right) \eqno {53} $$
with no logarithm.  The calculation of the fourth term gives only the
$\Delta$ term in the denominator and exhibits Hadamard behaviour .
\vskip .2truein
\noindent
{\bf{4.Conclusion}}

\noindent
We have already noted [10] that  while solving eq. [17]  after making the
ansatz
$\phi_1=\phi_0 (g(z,U)+h(\overline {z},U))$, we are solving an equation of 
the type
$$ [R{{\partial}\over {\partial U}}-{{(k_1+ik_2)}\over {\sqrt{2}}} 
{{\partial}\over {\partial z}}]g=I \eqno {54} $$
where $I$ denotes the inhomogeneous part of the equation.  This equation also
has a solution for the case when $I=0$, the homogeneous case.
Indeed, any arbitrary function
with the argument ${{U}\over {R}} (k_1+ik_2) + \sqrt {2} z $ is a solution.
If the function $f$ has a different dimension as compared to $\phi_0$,
then we have to multiply the solution to the homogenous equation by a 
dimensional constant.  In our problem the only such quantities are the modes,
$R$ and $k_1+ik_2$ that exist in  eq. [14].
This seems to be completely innocent, as far as $g$ is concerned.  The fact 
that we have
to sum over all the modes, however, changes  the result obtained for $G_F$.
One shows that we can generate $H_0^{(2)}$, the function that modifies the
Hadamard
behaviour in this way.  Which one of the two factors, $R$ or $k_1+ik_2$, is
used does not change the character of the new singularity structure.  It was
shown in  [10] that only $H_0^{(2)}$ survives the two derivatives
ifwe want to extract the vacuum expectation value of the stress-energy 
tensor, $<T_{\mu \nu}>$ from $G_F$ using the established methods $^{/2}$.
Itis also shown that this new singularity structure which already exists 
inthe homogeneous solution, and which is here unsurfaced by taking different
combinations for the inhomogenous term, does not change the result that
plane waves do not polarize the vacuum $^{/8,9}$.

Here we show another way to generate such spurious effects.
We refer to our earlier work
$^{/10}$ to establish that the renormalized value of the
VEV of the stress-energy tensor is independent of all these
spurious effects. The essence of the argument is that there is no finite part
of the resulting expression when the coincidence limit is taken, since
we have only an isolated pole divergence, and no remaining finite part.

We see that  there is more than one way to generate spurious singularities
for perturbatively computed Green functions.  Here  we study examples which
are obtained by grouping terms in different manners in inhomogenous
differential
equations.  It is a relief that these spurious effects do not change the
value calculated for $<T_{\mu \nu} >_{ren} $.

\vskip .2truein
\noindent
{\bf{Acknowledgement}}:  We are grateful to Prof.Dr. Yavuz Nutku for giving
us
his metrics and general support.  This work is partially supported by
T\" UB\.ITAK,
the Scientific and Technical Research Council of Turkey.  M.H.'s work is
also partially supported by T\" UBA, the Academy of Sciences of Turkey.

\vfill\eject
\noindent
{\bf{REFERENCES}}

\noindent
\item{1.} K.G.Wilson, Phys. Rev {\bf{D2}}, (1970) 1473 and 1478 ;

\item{2} N.D.Birrell and P.C.W. Davies, {\it{Quantum Fields in Curved Space
Space}}, Cambridge
University Press, Cambridge (1982) , S.A. Fulling, {\it{ Aspects of Quantum
Field Theory in Curved Space-Time}}
Cambridge University Press, Cambridge, 1989, R.M. Wald,{\it{ Quantum Field
Theory in Curved Spacetime and Black Hole Thermodynamics}}, The University of Chicago Press,
Chicago, 1994;

\item{3.} S.A.Fulling, M.Sweeney and R.Wald, Comm. Math. Phys.
{\bf{63}} (1978) 257;

\item{4.} J.Hadamard,{\it{Lectures on Cauchy's Problem in Linear Partial
Differential Equations}},
Yale University Press, New Haven (1923) ;

\item {5.} Earlier work can be found in M.Horta\c csu and R. Kaya ,
Gen. Rel. Grav. {\bf{30}} (1998) 1645;

\item{6.} P.A.Hogan, 1997 " Plane gravitational Waves and Bertotti-
Robinson Space-Times", University College Dublin , preprint;

\item{7.} Y.Nutku in {\it{Directions in General Relativity, Papers in honor
of Charles Misner}}
ed. by B.L.Hu, M.P.Ryan Jr. and C.V. Vishveshara,
Cambridge University Press , Cambridge (1993);

\item{8.} S.Deser, J. Phys. A: Math. Gen. {\bf{81}} (1975) 972;

\item{9.} G.W.Gibbons, Comm. Math. Phys. {\bf{45}} (1975) 191;

\item{10.} M.Horta\c csu and K.\" Ulker, , Class. Quantum Grav.,
{\bf{15}} (1998) 1415;

\item{11.} B.C.L\" utf\" uo\u glu, I.T.U. Thesis, unpublished.

\end